\documentclass[aps,prb,twocolumn,superscriptaddress,showpacs]{revtex4}
\usepackage{amsmath}
\usepackage{amssymb}
\usepackage{graphicx}
\usepackage{dcolumn}
\usepackage{bm}
\usepackage{placeins}

\begin{document}

\newcommand{\IKP}{Institut f{\"u}r Kernphysik, Technische Universit{\"a}t
Darmstadt, 64289 Darmstadt, Germany}
\newcommand{\YU}{Center for Theoretical Physics, Sloane Physics Laboratory, Yale University, New Haven, CT 06520-8120, USA}

\title{Lifshitz and Excited State Quantum Transitions in Microwave Dirac Billiards}

\author{B.~Dietz} \email{dietz@ikp.tu-darmstadt.de}
\affiliation{\IKP}
\author{F.~Iachello}
\affiliation{\YU}
\author{M.~Miski-Oglu}
\affiliation{\IKP}
\author{N.~Pietralla}
\affiliation{\IKP}
\author{A.~Richter}
\affiliation{\IKP}
\author{L.~von Smekal}
\affiliation{\IKP}
\author{J.~Wambach}
\affiliation{\IKP}

\date{\today}

\begin{abstract}
We present experimental results for the density of states (DOS) of a superconducting microwave Dirac billiard which serves as an idealized model for the electronic properties of graphene. The DOS exhibits two sharp peaks which evolve into Van Hove singularities with increasing system size. They divide the band structure into regions governed by the \emph{relativistic} Dirac equation and by the \emph{non-relativistic} Schr\"odinger equation, respectively. We demonstrate that in the thermodynamic limit a topological transition appears as a neck-disrupting Lifshitz transition in the number susceptibility and as an excited state transition in the electronic excitations. Furthermore, we recover the finite-size scaling typical for excited state quantum phase transitions involving logarithmic divergences and identify a quasi-order parameter.    
\end{abstract}


\maketitle
\section{\label{intr}Introduction}
\begin{figure}[!h]
\includegraphics[width=\linewidth]{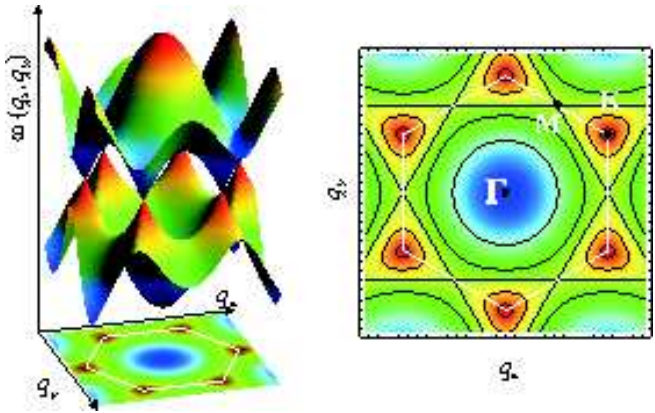}
\caption{(Color on line) The left panel shows the numerically determined conduction and valence band. They touch each other conically at the corners of the first Brillouin zone (white hexagon). The right panel shows the corresponding density plot in the quasi-momentum plane $(q_x,q_y)$ with the isofrequency lines shown as dark lines. The $\Gamma$ point denotes the maximum (minimum) of the conduction (valence) band, the $M$ points the saddle points and the $K$ points the touching points.} 
\label{fig1}
\end{figure}

Graphene, a monolayer of carbon atoms forming a hexagonal lattice, has attracted a lot of attention in recent years due to its extraordinary properties associated with the shapes of the conduction and the valence band shown in the left panel of Fig.~\ref{fig1}. These touch each other conically at the so-called $K$ or Dirac points, thus implying a linear dispersion relation. As a consequence, in their vicinity excitations are decribed by a Dirac Hamiltonian.~\cite{Semenoff1984} Indeed, even though the electrons move with a velocity which is 300 times smaller than the speed of light, graphene exhibits relativistic phenomena in the cone region.~\cite{Beenakker2008,Castro2009} Therefore, we refer to it as the \emph{relativistic} region.

Figure~\ref{fig1} shows in the left panel the band structure $\omega (\vec q)$ and in the right panel its
isofrequency lines~\cite{footnote1} (black lines) in the plane of the quasi-momentum vector components $(q_x,q_y)$.
The Dirac ($K$) points are located at the corners of the Brillouin zone (BZ). In their vicinity the isofrequency lines form circles that deform into triangles further away. This \emph{relativistic} region is bordered by saddle points at the $M$ points. At the centre of the Brillouin zone, the $\Gamma$ point, the conduction (valence) band has a maximum (minimum). In its vicinity the isofrequency lines form circles and the band structure has a parabolic shape. There, the Dirac equation is no longer applicable, i.e., the system is governed by the non-relativistic Schr\"odinger equation. Accordingly, the band structure can be separated into two independent \emph{relativistic} regions and a \emph{non-relativistic} one. A topologcal transition takes place at the $M$ points, where the Dirac cones merge into the parabolically shaped surface. There, due to a vanishing group velocity $\vert\vec\nabla \omega(\vec q)\vert =0$, the density of states $\rho$ (DOS) diverges logarithmically in an infinitely extended sheet of graphene.~\cite{Castro2009} These  "Van Hove singularities" (VHSs) have been predicted in general 2-dimensional crystals with a periodic structure.~\cite{Van1953} In bounded sheets the DOS exhibits peaks of finite height at the VHSs. 

We demonstrate that the topological transition at the $M$ points can be identified with a neck-disrupting ground-state Lifshitz transition.~\cite{Lifshitz1960} Such a transition has been observed experimentally only recently in two realizations of artificial graphene~\cite{Tarruell2012,Gomes2012} and in a microwave tight-binding analogue of graphene.~\cite{Bellec2013} There, a Lifshitz phase transition from a semimetallic to an insulating phase was induced with a controllable anisotropy in the honeycomb lattice.~\cite{Zhu2007,Montambaux2009} In Refs.~\cite{Son2011,Gradinar2012} Lifshitz transitions were investigated theoretically in sliding, respectively, strained bilayer graphene. We report on a gapless topological transition from the \emph{relativistic} to the \emph{non-relativistic} region induced by applying a chemical potential without changing the lattice structure.~\cite{Varlamov1989} In Ref.~\cite{McChesney2010} an experiment using angle-resolved photoemission spectroscopy was performed where the Fermi surface of graphene was gradually lifted to the VHS by chemical doping. However, the electron-electron and electron-phonon interactions have hampered the observation of the topological transition. Our microwave system, by construction, is free of such interaction effects. 

We will show that the topological transition can as well be associated with an excited state quantum phase transition (ESQPT) in the single-particle excitations~\cite{Caprio2008} as observed in the equivalent bosonic system and numerous other systems.~\cite{Iachello1996,Heiss2005,Leyvraz2005,Cejnar2006,Larese2011,Caprio2008} A particularly close analogy with the present case is provided by the 2-dimensional vibron model~\cite{Iachello1996} describing transverse vibrations of molecules. 

Lifshitz transitions and ESQPTs exhibit a characteristic scaling behavior of the "Van Hove" peak heights with the system size. For its experimental validation it is essential that the sharp peaks are not distorted by fermionic interactions. Thus, the scaling behavior cannot be determined through measurements in natural graphene~\cite{Kravets2010,Mak2011} where excitonic effects lead to a broadening and a shift of the peaks at the VHSs. Actually the phenomena associated with the band structure of graphene that we focus on are solely due to the presence of two interpenetrating triangular lattices with threefold rotational symmetry in the hexagonal lattice.~\cite{Wallace1947} Therefore, experiments with superconducting microwave Dirac billiards~\cite{Bittner2010,Bittner2012} are advantageous for the investigation of these phenomena since they correspond to idealized, non-interacting graphene. Another advantage, also encountered in "artificial graphene", where many-body effects are controllable (see, e.g., Ref.~\cite{Polini2013} for an overview) is that both systems can be taylored with a high degree of flexibility according to the phenomenon under investigation. 

\section{\label{exp}Experimental setup}
Superconducting microwave billiards have been used for two decades as analog systems for the study of non-relativistic quantum phenomena in high-resolution measurements.~\cite{Richter1999,StoeckmannBuch2000} Photonic crystals~\cite{Yablonovitch1989,Joannopoulos2008} are the optical analog of a solid and the frequencies of wave propagation as function of the two components of the quasi momentum exhibit a band structure. Both concepts can be combined into "microwave photonic crystals" which offer the opportunity to perform high-precision measurements of the excitation spectrum. The realization of a two-dimensional hexagonal structure utilizes metallic cylinders in a triangular lattice array~\cite{Smirnova2002,Bittner2010} squeezed between two metal plates. The structure of the first two frequency bands is similar to the band structure of graphene, that is, it is Dirac like in the vicinity of their touching points.~\cite{Raghu2008} Various effects have already been studied as, e.g., pseudo-diffusive transport near the Dirac point \cite{Zhang2008,Zandbergen2010,Bittner2012}, the quantum Hall effect \cite{Poo2011}, Zitterbewegung \cite{Zhang2008}, and edge states.~\cite{Zandbergen2010,Bittner2012,Kuhl2010a} 

\begin{figure}[h]
\includegraphics[width=\linewidth]{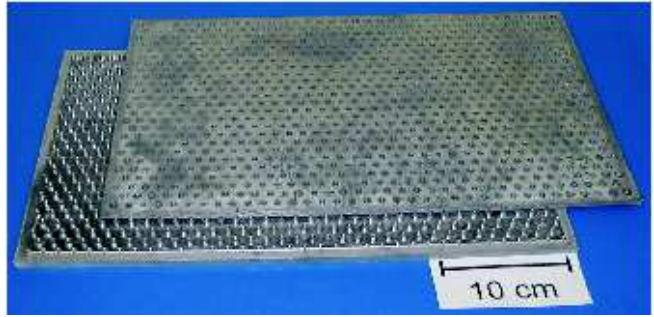}
\caption{(Color on line) Superconducting microwave Dirac billiard containing 888 metal cylinders. It is constructed from brass and coated with lead. The lid is shifted with respect to the billiard body.} 
\label{fig2}
\end{figure}

Here we present results associated with the properties of the DOS determined experimentally for two superconducting Dirac billiards.~\cite{Bittner2012} They consist of a brass lid and a rectangular brass basin with side lengths $420.0\times 249.4~{\rm mm}^2$ containing the metallic cylinders that are milled out of the plate. One Dirac billiard contained 267 cylinders and had the lattice constant $a_L={\rm 20 mm}$, the other one 888 with $a_L={\rm 12 mm}$. The radius of the cylinders was $R=a_L/4$. Figure \ref{fig2} displays the Dirac billiard with 888 cylinders, milled out of the bottom plate. 
The lids and the basins were lead coated to achieve superconductivity at liquid helium temperature. To ensure a good electrical contact the lids were screwed tightly to each cylinder. The height of the Dirac billiards was $h={\rm 3 mm}$. Hence, up to a maximum frequency of $50$~GHz, only the lowest transverse magnetic mode with the electric field vector perpendicular to the top and bottom plates was excited. Accordingly, the vectorial Helmholtz equation reduces to a scalar one which is mathematically identical to the Schr\"odinger equation of the corresponding 2-dimensionl quantum multiple-scattering problem with the waves scattered specularly at the walls of the cylinders and the billiard. 

\begin{figure}
\includegraphics[width=\linewidth]{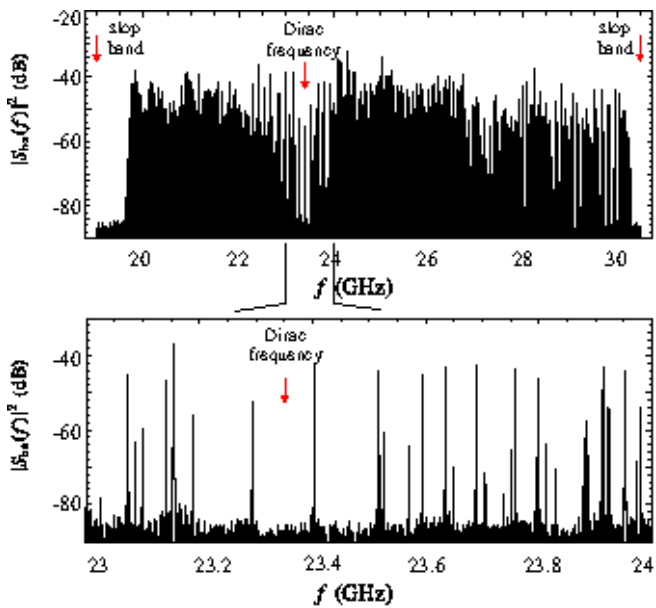}
\caption{High-resolution transmission spectrum of the microwave Dirac billiard depicted in Fig.~\ref{fig2} (upper panel). It is terminated by two stop bands, where no wave propagation is possible. The lower panel shows a zoom into the region of particularly low resonance density around the Dirac frequency.} 
\label{fig3}
\end{figure} 
For the measurement of the resonance spectra, the microwave power was coupled into and out of the resonator via wire antennas that reached a few millimeters into the resonator through holes in the lid. A Vector Network Analyzer measured the relative phase and amplitude of the output to the input signal. Transmission spectra were measured with all possible combinations of two out of a total of five antennas attached to the lid at different positions. Since the resonances had high quality factors $Q>{\rm5\cdot10^5}$, we could resolve all resonances and determined $1651$ eigenfrequencies. 

\section{\label{ExpDOS}Experimental resonance spectra and DOS}
In the upper panel of Fig.~\ref{fig3} a transmission spectrum of the Dirac billiard with 888 cylinders measured in the frequency region between 19.5~GHz and 30.5~GHz is depicted. It is bordered by two stop bands corresponding to the gaps in the band structure where no wave propagation is possible. Furthermore, we observe a region with an exceptionally low resonance density around the Dirac frequency of the Dirac points. The lower panel of Fig.~\ref{fig3} shows a zoom into it. 
\begin{figure}[!h]
\includegraphics[width=\linewidth]{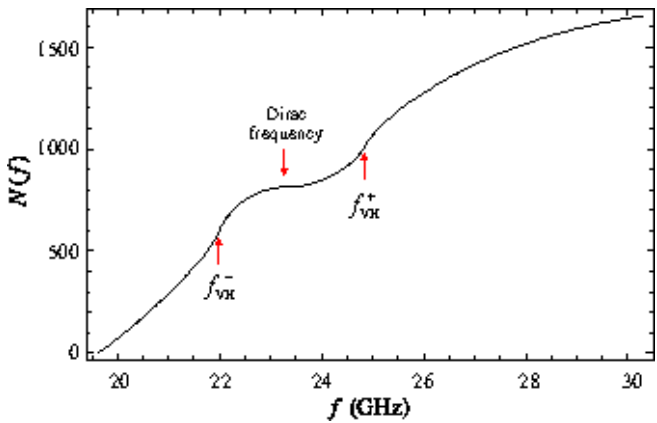}
\caption{(Color on line) The integrated resonance states $N(f)$ obtained from the resonance spectrum shown in Fig.~\ref{fig3}. It exhibits a plateau around the Dirac frequency $f_{\rm D}=23.36$~GHz, where it barely varies, and a slight kink at the frequencies denoted by $f^-_{\rm VH}=21.98$~GHz and $f^+_{\rm VH}=24.87$~GHz. Its frequency dependence below $f^-_{\rm VH}$ and above $f^+_{\rm VH}$ clearly differs from that inbetween.} 
\label{fig4}
\end{figure} 
\begin{figure}[!h]
\includegraphics[width=\linewidth]{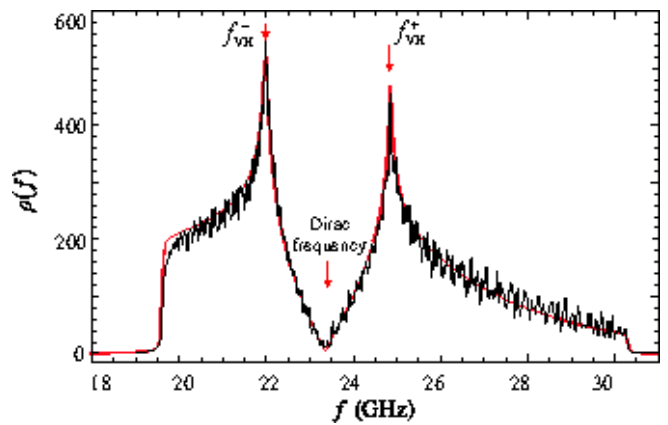}
\caption{(Color on line) Density of states $\rho (f)$ obtained from the resonance spectrum shown in Fig.~\ref{fig3} (black line). The red line results from a tight-binding model~\cite{Reich2002} (see text). The region of low density around the Dirac frequency $f_{\rm D}=23.36$~GHz is bracketed by two sharp peaks at $f^-_{\rm VH}=21.98$~GHz and $f^+_{\rm VH}=24.87$~GHz, which exhibit singular behavior at infinite system size.} 
\label{fig5}
\end{figure} 

Figure~\ref{fig4} shows the integrated resonance density $N(f)$ inferred from the measured resonance spectra as function of the excitation frequency $f$. As a consequence of the band structure of the photonic crystal inside the microwave billiard, $N(f)$ obviously differs from that of an empty one.~\cite{Richter1999,StoeckmannBuch2000} In a region around the Dirac frequency $f_{\rm D}=23.36$~GHz it exhibits a plateau reflecting the low density observed in that frequency range in the resonance spectrum (see Fig.~\ref{fig3}). Above (below) $f_{\rm D}$ it has the shape of half a parobola opening upwards (downwards). At the frequencies denoted by $f^-_{\rm VH}=21.98$~GHz and $f^+_{ \rm VH}=24.87$~GHz $N(f)$ has a slight kink. Below $f^-_{\rm VH}$ and above $f^+_{\rm VH}$ its frequency dependence is different from that in between. This is visible more clearly in the DOS shown in Fig.~\ref{fig5}. For its determination we counted the states $\Delta N(f)$ in frequency intervals $\Delta f=100$~MHz around $f$ and thus obtained $\rho(f)=\Delta N(f)/\Delta f$. The red curve has been computed from a tight-binding approach which incorporates not only the nearest-neighbor coupling $t$, but also the second- and third-nearest neighbor couplings $t_2$ and $t_3$ as well as the corresponding overlaps $s,\, s_2$ and $s_3$. Details on this model and on the definition of these quantities are given in Ref.~\cite{Reich2002}. A fit of the tight-binding model to the DOS yielded $(t=4.57,\, s=0.26)$, $(t_2=-0.28,\, s_2=-0.00001)$ and $(t_3=0.10,\, s_3=0.004)$. 

Around the Dirac frequency $f_{\rm D}$ the DOS vanishes linearly with $\vert f-f_{\rm D}\vert\to 0$. This region corresponds to the \emph{relativistic} one in the band structure, where the propagation of electromagnetic waves is governed by the Dirac equation.~\cite{Raghu2008} It is bracketed by two sharp peaks at $f^-_{\rm VH}$ and $f^+_{\rm VH}$. These are the VHSs.~\cite{Van1953} In the frequency range below $f^-_{\rm VH}$ and above $f^+_{\rm VH}$ the system is described by the Schr\"odinger equation of the corresponding quantum multiple-scattering problem. This defines the \emph{non-relativistic} region. A closer look at the experimental DOS reveals that the amplitudes and the typical frequencies of the oscillations of the experimental DOS are smaller in the frequency range between the two VHSs than below and above, thus indicating that both regions are governed by different wave equations. It should be noted that, to our knowledge, our measurement of the DOS including its fluctuations is the most precise so far. 

At the VHSs the DOS diverges logarithmically only for 2-dimensional structures of infinite extent. In the Dirac billiards used in the experiments, however, the sharp peaks at $f^{\pm}_{\rm VH}$ have a finite height $\rho^{\rm max}$. We determined it for the experimental DOS of the two microwave Dirac billiards, and also performed numerical studies for photonic crystals of various sizes with the shapes of rectangular and Africa billiards.~\cite{Berry1987} For a comparison of these results we rescaled the frequencies such that the distance $f^+_{\rm VH}-f^-_{\rm VH}$ between the VHSs, i.e., the group velocity, was the same for all systems. We chose the rescaling $f\rightarrow \tilde f$ such that $\tilde f^+_{\rm VH}-\tilde f^-_{\rm VH}=2$. 
The experimental and numerical studies revealed that the maxima of the DOS, $\rho^{\rm max}$, or rather those of the renormalized DOS, $n^{\rm max}=\frac{f^+_{\rm VH}-f^-_{\rm VH}}{2}\rho^{\rm max}$, behave like
\begin{equation}
n^{\rm max}\simeq a N_c\left(\ln(N_c)+b\right)\label{rhoexp}
\end{equation}
with $N_c$ the number of unit cells, i.e., of hexagons formed by the voids in the photonic crystal. The quantities $a$ and $b$ are fit parameters. The latter depends on the size of the frequency interval $\Delta f$ chosen for the computation of $\rho (f)=\Delta N/\Delta f$, while the former takes a similar value $a\sim 0.145-0.155$ for all cases, i.e., it seems to be \emph{universal}. This finite-size scaling, which is also typical for an ESQPT~\cite{Caprio2008}, and the fate of the isofrequency lines at the saddle points (see right panel of Fig.~\ref{fig1}) suggests a description in terms of a neck-disrupting Lifshitz transition.~\cite{Lifshitz1960} We should note that all properties of the DOS that we observe coincide with those of the DOS for vibrations perpendicular to the plane of an hexagonal lattice, as shown by Hobson and Nierenberg.~\cite{Hobson1953} 

\section{\label{Lifshitz}Neck-disrupting Lifshitz transition}
In order to illustrate the relation between the VHSs in the DOS of the microwave photonic crystal and the neck-disrupting Lifshitz transition
in the corresponding fermionic band structure, we have computed the number susceptibility from the particle-hole polarization (Lindhard) function.~\cite{Lindhard1954} For this we used the simplest tight-binding model, which takes into account only nearest-neighbor hopping of strength $t$.~\cite{Castro2009,Reich2002} Many aspects concerning the electronic excitations in graphene at weak coupling~\cite{Castro2009} have been studied analytically with this model primarily in the Dirac cone approximation, to exemplify more general effects.~\cite{footnote2} Details concerning the computation of the retarded particle-hole polarization function $\Pi(\omega,\vec p; \mu)$, with $\omega$ the excitation frequency, $\vec p$ the vector of momentum transfer and $\mu$ the chemical potential~\cite{Stauber2010,Kotov2012} are given in the appendix. It is a sum of particle-hole transitions within the same band, $\Pi^+$, i.e., \textit{intraband} transitions and those arising from \textit{interband} transitions between the two bands, $\Pi^-$. 

Static Lindhard screening is described by the retarded susceptibility
\begin{equation}
\chi^R(\vec p) =  \Pi(\omega=0, \vec p; \mu) \; .
\end{equation}
The usual Thomas-Fermi susceptibility is in turn defined as the subsequent long-wavelength limit $\chi = \lim_{\vec p\to 0} \chi^R(\vec p)$. The imaginary part of $\Pi(\omega=0, \vec p; \mu)$ vanishes in the static limit as long as the spatial momentum is nonzero. In the long-wavelength limit, on the other hand, only interband transitions survive. These yield for the (zero-temperature) number susceptibility~\cite{Stauber2010}
\begin{equation} 
\begin{split}
\chi&= \lim_{\vec p\to 0}\lim_{\omega\to 0}  \Pi(\omega,\vec p;\mu)=\frac{\rho(\mu)}{A}.   
\end{split} \label{chiisdos}
\end{equation} 
Hence, it coincides with the DOS $\rho (\omega)$ per area $A$ of the graphene sheet at the Fermi surface $\omega =\mu$. 

Adapting the definitions of the frequency scale from Ref.~\cite{Castro2009}, the zero of the DOS, identified with the Dirac point, is located at $\mu=0$, the VHSs are at $\mu=\pm t$ and the band gaps start at $\mu=\pm 3t$. When the chemical potential is chosen near one of the VHSs we readily obtain from the analytical expression Eq.~(14) in Ref.~\cite{Castro2009} for the fermionic system at finite-charge density
\begin{equation}
\rho(\mu)  = \frac{3N_c}{2\pi^2A t} \Big\{  - \frac{1}{2}
\ln\Big(\frac{|\mu |}{t} - 1\Big)^2 + 2\ln 2 + \mathcal
O\Big(\frac{|\mu|}{t} -1\Big) \Big\}  \, . \label{vHs}
\end{equation}
The divergence of  $\chi$ as $\vert\mu\vert\to t$ is caused by the infinite degeneracy of ground states of the 2-dimensional system when the Fermi surface passes through a VHS. In the thermodynamic sense this can be considered as a zero-temperature quantum phase transition with control parameter $|\mu|$. To illustrate this we introduce the reduced Fermi-energy parameter $z
= (|\mu|-t)/t$ to rewrite Eq.~(\ref{vHs}),
\begin{equation}
\chi(z)  = \frac{3N_c}{2\pi^2A t} \Big(  -
\ln |z|  + 2\ln 2 + \mathcal O(z) \Big)  \, . \label{LT}
\end{equation}
Unlike the cases of first or second order phase transitions, the susceptibility does not diverge with a power law in $z$ but logarithmically. This is a manifestation of the neck-disrupting Lifshitz transition in two dimensions.~\cite{Lifshitz1960,Blanter1994} The singular part of the corresponding thermodynamic grand potential is non-zero on both sides of the transition. Following Ref.~\cite{Blanter1994}, it is given per area of the sample $A$ by    
\begin{equation}
\frac{\Omega_\mathrm{sing}}{A} = \frac{3N_c}{2\pi^2A t} \Big( \frac{(tz)^2}{2} +
\frac{\pi^2}{6}T^2\Big) \, \ln|z| \, .
\end{equation}

The susceptibility or DOS does not diverge in a 2-dimensional system of finite area {\it A}. To see how the heights of its maxima scale with  {\it A} we used periodic boundary conditions and integrated Eq.~(\ref{LT}) over a small interval $\Delta z =(2\pi)^2/N_c$ around the singularity. After rescaling the energies such that the distance between the maxima equals $2$ we obtain for the height of the maxima of the renormalized DOS $n^\mathrm{max}=t\rho^\mathrm{max}$
\begin{equation}
  n^\mathrm{max}\simeq\frac{3}{2\pi^2} \, N_c\, \Big(\ln N_c - 2\ln\pi
  +1  \, + \mathcal{O} (1/N_c) \Big) \; . \label{fssLT}
 \end{equation} 
Note that $\frac{3}{2\pi^2}\simeq 0.15$, thus confirming the experimental and the numerical findings, c.f., Eq.~(\ref{rhoexp}). Thus the height of the maxima of the susceptibility at the VHSs scales as $t\chi^\mathrm{max} = n^\mathrm{max}/A \sim \ln N_c $, in accordance with the finite-size scaling of a neck-disrupting Lifshitz transition. The transition is due to a change of topology of the Fermi surface with no order parameter in the strict sense. We present a quasi-order parameter below.

\section{\label{ESQPT}Excited state quantum transition in the electronic excitations} 
\begin{figure}
\includegraphics[width=\linewidth]{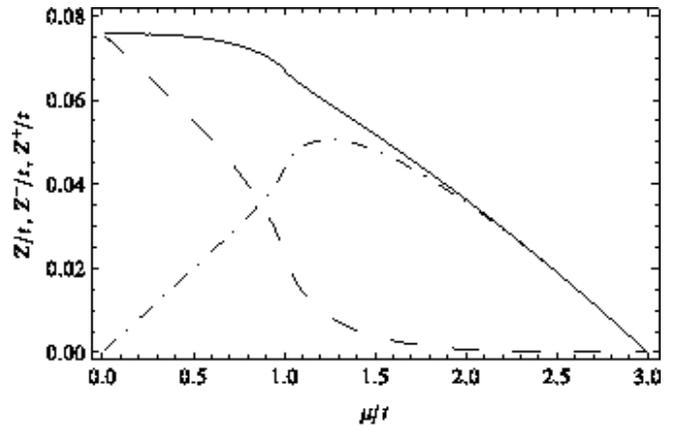}
\caption{$f$-sum rule evaluated separately for interband (dashed line) and intraband (dashed-dotted line) transitions and their sum (full line). For $\mu/t=1$, i.e., when the Fermi surface passes through a VHS, all three curves exhibit a rapid change. There, their derivatives are logarithmically divergent.}
\label{fig6}
\end{figure}

The singularity of the single-particle DOS as function of the excitation frequency also shows up in the spectrum of particle-hole excitations. This is reminiscent of the ESQPT observed for the vibrational modes of molecules.~\cite{Iachello1996,Caprio2008} Clear support for an interpretation as an ESQPT is provided by the universal finite-size scaling behavior (Eq.~(\ref{rhoexp})) typical for it. To further quantify the analogy,  we analyze the polarization function at zero-momentum transfer, $\Pi(\omega,\vec p=0; \mu)$. The associated spectral distribution $\rho_\mathrm{ph}(\omega)$ of particle-hole excitations is given by
\begin{equation}      
\rho_\mathrm{ph}(\omega) = Z(\mu)^{-1} \lim_{\vec p^2  \to 0} \, \frac{\omega}{2\pi \vec p^2} \,
\mbox{Im} \, \Pi(\omega,\vec p; \mu) \; .
\label{rhodef}
\end{equation}
The normalization 
\begin{equation} 
Z(\mu) =  \lim_{\vec p^2  \to 0} \,\int_0^\infty
d\omega  \frac{\omega}{2\pi\vec p^2} \,
\mbox{Im} \, \Pi(\omega,\vec p; \mu) \; .       \label{norm}
\end{equation}
can be separated into contributions $Z^+$ from intraband and $Z^-$ from interband transitions, respectively. For the latter, analytic results only exist in the Dirac cone approximation. We have extended this appropriately and include the results in the appendix, see Eqs.~(\ref{Zp}) and ~(\ref{Zm}). Figure~\ref{fig6} depicts the intraband $Z^+$ and interband $Z^-$ contributions and their sum $Z=Z^++Z^-$. 
The intraband term $Z^+$ is fixed due to charge conservation via the $f$-sum rule~\cite{Nozieres1997,Sabio2008,Kotov2012} in terms of the 2-dimensional charge carrier density $n_c$ and mass $m$ as  $Z^+=\frac{n_c}{4 m}$. Near the centre of the Brillouin zone we have $n_c=p_F^2/(4\pi)$ and $m=\sqrt{3}N_c/(tA)$ and the contribution from interband transitions behaves as $Z^-(\mu)\simeq\frac{1}{108}\frac{1}{8\pi}(3t - |\mu|)^3$. Hence it is suppressed with respect to $Z^+$ such that $Z(\mu)\approx Z^+(\mu)$, and the sum rule is readily verified, 
\begin{equation}
 Z(\mu)\simeq  \frac{1}{8\pi} \, (3 t - |\mu|)  = \frac{1}{8\pi}
 \,  \frac{p_F^2}{2m} = \frac{n_c}{4 m}\, . 
\end{equation}
This approximation holds in the non-relativistic Fermi liquid regime either below or above the two VHSs,
i.e., for $|\mu| > t$. Near the Dirac cone, where $|\mu| \ll t$, on the
other hand, the intraband transitions yield    
\begin{equation}
Z^+(\mu)\simeq
\frac{|\mu|}{8\pi}  = \sqrt{\frac{n_c'}{2\pi}} \, \frac{v_F}{4}  \, ,\label{ZplusDC}
\end{equation}  
with $n_c' =\mu^2/(2\pi v_F^2)  $. Thus, there the intraband $f$-sum rule 
scales with the square root of the carrier density $n_c'$ relative to half
filling. However, the contribution of the interband transitions to $Z(\mu)$, 
\begin{equation}
Z^-(\mu)\simeq\frac{\pi t}{24\sqrt{3}} -\frac{|\mu|}{8\pi}\, ,\label{ZminusDC}
\end{equation}  
can no longer be neglected for $|\mu| \ll t$. Note that the sum $Z(\mu)$ of the contributions Eqs.~(\ref{ZplusDC}) and (\ref{ZminusDC}), is independent of $\mu$ and hence of the carrier density.~\cite{Sabio2008,Kotov2012} 

From these observations we conclude that the $f$-sum rule or $Z(\mu)$ 
can serve as a quasi-order parameter for the Lifshitz transition, indicating
relativistic behavior for $|\mu |/t < 1$
with $Z(\mu)\approx$ const., as compared to the non-relativistic Fermi-liquid regime for $|\mu|>t$, where $Z(\mu)$ decreases almost linearly with $\mu$. We verified analytically that the derivative of $Z(\mu)$ with respect to $\mu$ given in Eqs.~(\ref{derivIntra}) and (\ref{derivInter}) diverges logarithmically at the Lifshitz transition, $\mu=t$. This reflects a singular behavior of the carrier density similar to that of $\chi$ in Eq.~(\ref{LT}), since $Z^+\propto n_c$ for $|\mu|/t\gtrsim 1$. 

We obtained the full spectral distribution $\rho_\mathrm{ph}(\omega) $ of
particle-hole excitations Eq.~(\ref{rhodef}) from explicit analytical expressions 
for the polarization 
function. The results are given in Eqs.~(\ref{stauber}) and (\ref{EEDOS}). that for $\Pi^+(\omega,\vec p; \mu) $ was first derived in Ref.~\cite{Stauber2010}. For $\vec p\to 0$ and $\omega >0$ only the imaginary part of $\Pi^-(\omega,\vec p; \mu) $ is nonvanishing.
 \begin{figure}[!h]
\includegraphics[width=\linewidth]{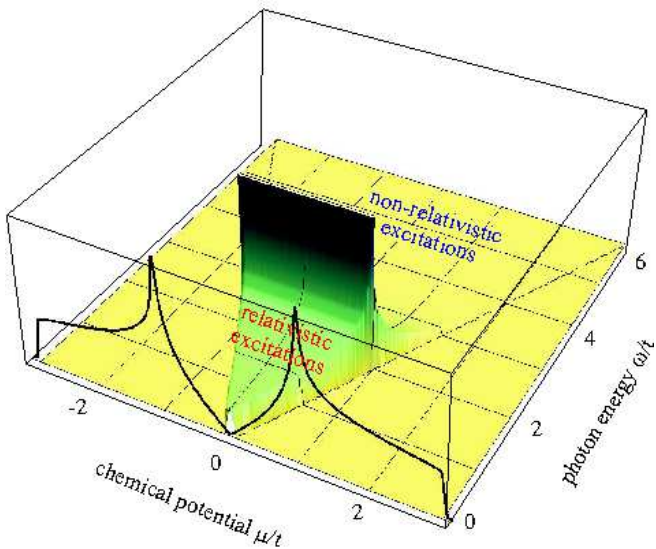}
\caption{(Color on line) Spectral distribution
  $\rho_\mathrm{ph}  (\omega) $ of vertical particle-hole excitations
  computed from Eq.~(\ref{rhodef}) as function of the rescaled chemical potential $\mu/t$ and excitation frequency $\omega/t$. It exhibits a logarithmic singularity at $\omega/t= 2 $ for $|\mu|/t < 1 $. Also
  displayed in the front panel for $\omega = 0$ is the number susceptibilty $\chi = \rho(\mu)/A$ to indicate the ground-state Lifshitz transitions at $\mu = \pm t$.}
\label{fig7}
\end{figure}
The result is illustrated in Fig.~\ref{fig7} where we display $\rho_\mathrm{ph}(\omega)$. In the front panel we have included the number susceptibility 
$\chi(\mu) = \rho(\mu)/A$, to indicate the two ground-state
Lifshitz transitions at $\mu = \pm t$. As in the Dirac
cone approximation, interband contributions to  $\rho_\mathrm{ph}(\omega)$ vanish when $\omega < 2\mu$ because the vertical particle-hole excitations
are then Pauli blocked. At $\omega =2t$ the spectral distribution exhibits a clearly visible divergence which is directly related to that of the single-particle DOS at the VHS (Eq.~(\ref{chiisdos})) as demonstrated in the appendix in Eq.~(\ref{EEDOS}). There, transitions take place between the saddle points of the valence and the conduction band, where the DOS is singular. Below this, for $\omega < 2 t$ and $\mu < t$, we have relativistic behavior of the low-frequency excitations. For $\omega > 2 t$, the density of particle-hole excitations decreases fast with increasing frequency, as it does in the normal Fermi-liquid regime. We associate the logarithmic singularity at $\omega =2t$ with an ESQPT from the relativistic region for $\omega < 2 t$ to the non-relativistic one for $\omega > 2 t$. Similarly, in molecules~\cite{Iachello1996,Caprio2008} the ESQPT becomes manifest in a logarithmic singularity of the level density. However, in distinction to our case, it is characterized by an order parameter. 

In Ref.~\cite{Mak2011} the optical conductivity of graphene was measured, which is related to the spectral distribution shown in Fig.~\ref{fig7} at $\mu /t=0$. 
Due to excitonic effects resulting from the electron-hole interactions a broadened peak was observed at the interband transition from the lower to the higher $M$ point, which was in addition shifted with respect to its predicted position. This peak in fact is a remnant of the ESQPT, which is hidden due to the many-body correlations in natural graphene. The measurement of a pure ESQPT, i.e., an experimental mapping of Fig.~\ref{fig7} should be possible with artificial graphene, since there the Fermi surface can be shifted via doping and at the same time the electron-hole interactions can be turned off.~\cite{Tarruell2012,Gomes2012}

\section{Conclusions}
We have determined the DOS in high-precision experiments with two superconducting Dirac billiards. It is similar to that of transverse vibrations of an hexagonal lattice and, most importantly, to that of the electronic band structure of finite sheets of graphene in the absence of fermionic interactions. In a second part we have shown that the properties of the observed DOS can be quantitatively related to a ground-state QPT and an ESQPT arising from the topological Lifshitz neck-disrupting phase transition. Due to the unprecedented accuracy in the determination of the DOS we were able to first recover the finite-size scaling governing such transitions. Furthermore we found a quasi-order parameter for the Lifshitz transition. An experimental verification of our analytical result for the spectral distribution shown in Fig.~\ref{fig7} should be possible with artificial graphene.~\cite{Tarruell2012,Gomes2012}

\section{Acknowledgements}
This work has been supported by the DFG within the SFB 634. F.I. acknowledges support from U.S.D.O.E. Grant DE-FG02-91ER40608, and L.v.S. from the European Commission, FP7-PEOPLE-2009-RG, No. 249203.

\appendix
\section{Computation of the Lindhard function}
We consider the nearest-neighbor tight-binding model without overlap
corrections. We will furthermore neglect the
physical spin of the electrons which would simply amount to a doubling
of the degrees of freedom here. The tight-binding
Hamiltonian is then given by the nearest-neighbor $\langle i,
j\rangle$ sum of hopping terms with strength $t$,
\begin{equation} 
\hat H = -t \sum_{\langle i,j\rangle} \big( a_i^\dagger b_j +
b^\dagger_j a_i \big) \; , \label{tight-binding}
\end{equation}
where $a^\dagger, a$ and $b^\dagger, b$ are the fermionic creation
and annihilation operators of the two distinct triagonal sublattices
that make the honeycomb lattice. It is readily diagonalised in momentum space
\cite{Wallace1947} where it is expressed in terms of the matrix
\begin{equation}
 H(\vec k) = - \big(\mu +  B_1(\vec k)  \, \sigma_1 + B_2(\vec k) \,
 \sigma_2  \big) \, . 
\end{equation}
Here we have included a chemical potential $\mu$ for a finite
charge-carrier density which for half filling equals $\mu =0$, $\sigma_i$ are Pauli matrices which act in
the space of the two sublattices, and $B_i(\vec k)$, $i=1,2$, are the real and
imaginary parts of the complex structure factor,
\begin{eqnarray} 
&&t\Phi(\vec k)=t \sum_{n=1}^3  e^{ i \vec k\cdot \vec\delta_n}\equiv B_1(\vec k) + i B_2(\vec k) \, ,\\ 
&&\nonumber\vec\delta_1=\frac{a_c}{2}(-1,\sqrt{3}),\, \vec\delta_2=\frac{a_c}{2}(-1,-\sqrt{3}),\, \vec\delta_3=a_c(1,0).~~~
\end{eqnarray}
with the nearest-neighbor vectors $\vec\delta_n$ on the honeycomb lattice and $a_c$ the carbon-carbon distance in graphene.

Using Dirac matrices $\gamma^0 \equiv \sigma_3 
$ and $\gamma^i \equiv \sigma_3\sigma_i$, with $i=1, 2$, for the
Clifford algebra  $\{\gamma_\mu,\gamma_\nu\} = 2 g_{\mu\nu} $ in the
two-dimensional sublattice space, we can introduce the free
fermion propagator on the honeycomb lattice as the resolvent
\begin{equation} 
\big(\gamma^0 (\omega - H)\big)^{-1} = \frac{\gamma^0 (\omega+\mu) - 
  \vec B \cdot \vec\gamma}{(\omega+\mu)^2 - E^2(\vec k)} \equiv
-iG_F(\omega,\vec k)\, ,
\end{equation}
where the roots of $ E^2(\vec k)  \equiv t^2 |\Phi(\vec k)|^2$
are the two single-particle energy bands $  E_\lambda(\vec k) = \lambda t |\Phi(\vec k)| $ of the model. They are given by~\cite{Hobson1953}
\begin{equation}
tE_\lambda(\vec k)=t\lambda\vert\Phi_{\vec k}\vert=t\lambda\sqrt{1+4F(\vec k)},
\end{equation}
\begin{equation}
F(\vec k)=\cos\left(\frac{\sqrt{3}}{2}k_y{a_c}\right)\cos\left(\frac{3}{2}k_xa_c)\right) +\cos^2\left(\frac{\sqrt{3}}{2}k_ya_c\right),\nonumber
\label{Energieflaeche}
\end{equation}
where $k_x$ and $k_y$ are the two components of the quasi momentum which is restricted to the first Brillouin zone (BZ), $\lambda=+1$ labels the conduction and $\lambda=-1$ the valence band. The corners of the BZ are at the $K$ points. The associated energies equal $E_\lambda(\vec k_{\rm D})=0$. At the saddle points, the so-called $M$ points, the energies are $E_\lambda(\vec k_M)=\lambda$. For $\lambda=1$ ($\lambda =-1$) the band structure has a maximum (minimum) at $\vec k_\Gamma=\vec 0$. These are the so-called $\Gamma$ points, which are located at the center of the BZ. The associated energies equal $E_\lambda(\vec k_\Gamma)=3\lambda$. 

The charge-density correlations are determined by the diagonal
time component of the corresponding polarisation tensor as in QED.
In the random-phase approximation (RPA) this particle-hole
polarisation function is given by the one-loop expression
\begin{eqnarray}
&&\Pi(\omega,\vec p;\mu)\\ 
=&&\int_\mathrm{BZ}\frac{d^2q}{(2\pi)^2}\int\frac{dq^0}{2\pi}
\mbox{tr}\Big(\gamma^0 G_F(q^0,\vec q)\gamma^0G_F(q^0+\omega,\vec q+\vec p)\Big)\, ,\nonumber
\end{eqnarray}
where the spatial loop momentum is integrated over the first Brillouin
zone (BZ). In the imaginary-time formalism, the sum over the Matsubara
frequencies $q^0 = i (2n+1)\pi T  $ in this RPA polarisation loop
can be evaluated for discrete $\omega = i 2\pi m T$ (integer
$m,n$) at finite temperature $T$ with standard techniques. After
analytic continuation back to real frequencies $\omega$, with retarded
boundary conditions say, this yields the Lindhard function~\cite{Lindhard1954} of the
honeycomb lattice in the form \cite{Stauber2010},
\begin{eqnarray}
\Pi(\vec q,\omega;\mu)&=&-\frac{1}{t}\frac{1}{2\pi^2}\int_{1. BZ}{\rm d}^2k\sum_{\lambda, \lambda^\prime=\pm 1}f_{\lambda\cdot\lambda^\prime}(\vec k,\vec q)\nonumber\\
\nonumber&\times&\frac{\left[n_F\left(E_{\lambda^\prime}(\vec k+\vec q)-\mu\right)-n_F\left(E_\lambda(\vec k)-\mu\right)\right]}{E_{\lambda^\prime}(\vec k+\vec q)-E_\lambda(\vec k)-\frac{\hbar\omega}{t}-i\epsilon}.\\&&
\label{Lindhard}
\end{eqnarray}
Here, the function
\begin{equation}
f_{\lambda\cdot\lambda^\prime}(\vec k,\vec q)=\frac{1}{2}\left(1+\lambda\cdot\lambda^\prime{\rm Re}\left[
\frac{\Phi_{\vec k}\Phi^\star_{\vec k+\vec q}}{\vert\Phi_{\vec k}\vert\vert\Phi_{\vec k+\vec q}\vert}\right]\right)
\label{f}
\end{equation}
takes account of the overlap between the wave functions associated with the two bands. 
For $\vec q=\vec 0$ we have $f_{\lambda\cdot\lambda^\prime}(\vec k,\vec q)=\frac{1}{2}\left(1+\lambda\cdot\lambda^\prime\right)$. Furthermore, $n_F(E)=\left(e^{E/T}+1\right)^{-1}$ is the Dirac-Fermi distribution. Since we are mainly interested in the zero temperature transition we set $T=0$, so $n_F(E)=\Theta(-E)$ equals the staircase function, and consider only excitations, i.e., we assume that $\omega\geq 0$. Then, either $\lambda^\prime =\lambda =1$, which corresponds to an intraband transition or $\lambda^\prime =-\lambda =1$ for interband transitions. The Dirac energy equals $\omega _q=tv_F\vert\vec q\vert$ with the Fermi velocity $v_F$ given in units of $t$ as $\frac{3}{2}a_c/\hbar$.

We computed the imaginary part of the Lindhard function which can be interpreted as the density of particle-hole excitations, 
\begin{eqnarray}
{\rm Im}\Pi(\vec q,\omega;\mu)&=&-\frac{1}{t}\frac{1}{2\pi}\int_{1. BZ}{\rm d}^2k\sum_{\lambda, \lambda^\prime=\pm 1}f_{\lambda\cdot\lambda^\prime}(\vec k,\vec q)\nonumber\\
\nonumber&\times&\left[\Theta\left(\mu-E_{\lambda^\prime}(\vec k+\vec q)\right)-\Theta\left(\mu-E_\lambda(\vec k)\right)\right]\\
\nonumber&\times&\delta\left[E_{\lambda^\prime}(\vec k+\vec q)-E_\lambda(\vec k)-\frac{\hbar\omega}{t}\right].\\&&
\label{ImLindhard}
\end{eqnarray}
Our main focus was its evaluation in the limit of small momentum transfers $\vec q$, i.e. $\vert\vec q\vert\to 0$. Thus, without loss of generality we may choose $\vec q$ in $\Gamma M$ direction, i.e., set $\vec q=(q_x,0)$. The integration over $k_x$ can be performed. For this we replace $\frac{3a_c}{2}k_x$, $\frac{\sqrt{3}a_c}{2}k_y$, $\frac{3a_c}{2}q_x$, $\frac{\hbar\omega}{t}$ and $\frac{\hbar\omega_q}{t}$ by, respectively, the dimensionless quantities $k_x$, $k_y$, $q_x$, $\omega$ and $\omega_q=q_x$ and define $y=\cos(k_y)$. Furthermore, we set $t$ and $a_c$ equal to 1 and we introduce $j=\pm 1$ and define $q_x=s\tilde q_x$ with $s=\pm 1$ such that the integral over the BZ is transformed to an integral over $0\leq k_x\leq \pi/2$ and $0\leq k_y\leq\pi /2$.~\cite{Stauber2010} With the notations
\begin{eqnarray}
&&a=2jy\cos\left(k_x+\frac{q_x}{2}\right)\\
\nonumber &&b=2jy\sin\left(k_x+\frac{q_x}{2}\right)
\end{eqnarray}
we obtain
\begin{eqnarray}
&&\nonumber\vert\Phi_{\vec k}\vert=\sqrt{1+4y^2+2(a\cos\frac{q_x}{2}+b\sin\frac{q_x}{2})}\\
&&\nonumber\vert\Phi_{\vec k+\vec q}\vert=\sqrt{1+4y^2+2(a\cos\frac{q_x}{2}-b\sin\frac{q_x}{2})}\\
&&\nonumber{\rm Re}\left[\Phi_{\vec k}\Phi^\star_{\vec k+\vec q}\right]=\cos\frac{2q_x}{3}+4y^2\cos\frac{q_x}{3}+2a\cos\frac{q_x}{6}.~~~~~~~
\\&&
\end{eqnarray}
Introducing the notations
\begin{equation}
\omega_q=2\sin\frac{q_x}{2},\, x=\left(\frac{w}{w_q}\right)^2,\, l=\pm 1\\
\end{equation}
the evaluation of the $\delta$-function yields 
\begin{eqnarray}
\omega&=&\vert\Phi_{\vec k+\vec q}\vert-\lambda\cdot\lambda^\prime\vert\Phi_{\vec k}\vert\\
a&=&a_l=-x\cos\frac{q_x}{2}+l\sqrt{\left(1-x\right)\left(4y^2-x\right)}\\
\vert b\vert&=&b_l=\sqrt{4y^2-a_l^2}\\
\label{Bl}B_l&=&\frac{\omega_q}{\omega}b_{l}=\sqrt{1+4y^2+2\cos\frac{q_x}{2}a_l-\frac{\omega^2}{4}}\\
&&\vert\Phi_{\vec k}\vert\vert\Phi_{\vec k+\vec q}\vert=\lambda\cdot\lambda^\prime\left(B_l^2-\frac{\omega^2}{4}\right)
\end{eqnarray}
with the requirements
\begin{eqnarray}
&&\label{requir}\vert a_l\vert\leq 2y\\
&&\nonumber\left(1-x\right)\left(4y^2-x\right)\geq 0\\
&&\nonumber\lambda\cdot\lambda^\prime\frac{\omega}{2}\leq\lambda\cdot\lambda^\prime B_l\, .
\end{eqnarray}
Furthermore we obtain $\vert\Phi_{\vec k}\vert=-\lambda\cdot\lambda^\prime\frac{\omega}{2}\pm B_l$ and thus for interband transitions ($\lambda\cdot\lambda^\prime=-1$)
\begin{eqnarray}
&&\omega=\left(\vert\Phi_{\vec k+\vec q}\vert +\vert\Phi_{\vec k}\vert\right)\\
&&\nonumber\vert\Phi_{\vec k}\vert =\frac{\omega}{2}\pm B_l\\
&&\nonumber\vert\Phi_{\vec k+\vec q}\vert =\frac{\omega}{2}\mp B_l\\
&&\nonumber\frac{\omega}{2}\geq B_l\, ,
\end{eqnarray}
and for intraband transitions ($\lambda\cdot\lambda^\prime=1$)
\begin{eqnarray}
&&\omega=\left(\vert\Phi_{\vec k+\vec q}\vert -\vert\Phi_{\vec k}\vert\right)\\
&&\nonumber\vert\Phi_{\vec k}\vert =-\frac{\omega}{2} +B_l\\
&&\nonumber\vert\Phi_{\vec k+\vec q}\vert =\frac{\omega}{2} +B_l\\
&&\nonumber\frac{\omega}{2}\leq B_l\, .
\end{eqnarray}
Using the property of the $\delta$-function
\begin{equation}
\label{deltapr}\delta (f(x))=\sum_i\frac{1}{\vert df(x)/dx\vert_{x=x_i}}\delta (x-x_i)\, ,\, f(x_i)=0
\end{equation}
we finally obtain for the imaginary part of the Lindhard function Eq.~(\ref{ImLindhard})
\begin{eqnarray}
\nonumber {\rm Im}\Pi(\vec q,\omega;\mu)=\frac{\sqrt{3}}{\pi}\frac{1}{\left(\hbar v_F\right)^2}\frac{1}{2\omega_q}\int_{y_{min}}^{y_{max}}\frac{{\rm d}y}{\sqrt{1-y^2}}&&\label{Integrand}\\
\nonumber\times\frac{\Theta\left(\lambda\cdot\lambda^\prime\left[B_l-\frac{\omega}{2}\right]\right)}{\sqrt{\left(1-x\right)\left(4y^2-x\right)}}\left[\frac{T^+_{\lambda\cdot\lambda^\prime}F^+_{\lambda\cdot\lambda^\prime}}{B_+}+\frac{T^-_{\lambda\cdot\lambda^\prime}F^-_{\lambda\cdot\lambda^\prime}}{B_-}\right]&&\\&&
\end{eqnarray}
where $v_F=\frac{3a_c}{2\hbar}$ is the Fermi velocity in units of $t$ and 
\begin{eqnarray}
\label{Tp1}T^l_{\lambda\cdot\lambda^\prime=1}&=&
\Theta\left(\frac{\omega}{2}+B_l-\mu\right)-\Theta\left(-\frac{\omega}{2}+B_l-\mu\right)~~~~~~~\\
\label{Tm1}T^l_{\lambda\cdot\lambda^\prime=-1}&=&
\Theta\left(\frac{\omega}{2}+B_l-\mu\right)+\Theta\left(\frac{\omega}{2}-B_l-\mu\right)~~~~~~
\end{eqnarray}
and
\begin{equation}
\lambda\cdot\lambda^\prime F^l_{\lambda\cdot\lambda^\prime}=B^2_l-\frac{\omega^2}{4}
+\left[\cos\frac{2q_x}{3}+4y^2\cos\frac{q_x}{3}+2a_l\cos\frac{q_x}{6}\right].
\end{equation}
The integration limits $y_{min}$ and $y_{max}$ are determined with the help of the requirements Eq.~(\ref{requir}).
\subsection{\label{ImP+}Intraband transitions for small momentum transfer}
In the limit $\omega_q\to 0$ the intraband transitions give a nonvanishing contribution to the integral Eq.~(\ref{ImLindhard}) only for $x=\left(\frac{\omega}{\omega_q}\right)^2<1$, i.e., for small excitation energies $\omega\leq\omega_q$. For $\omega\to 0$ the difference of the $\Theta$-functions Eq.~(\ref{Tp1}) can be approximated as 
\begin{equation}
\frac{\Theta\left(\frac{\omega}{2}+B_l-\mu\right)-\Theta\left(-\frac{\omega}{2}+B_l-\mu\right)}{\omega}\simeq \delta\left(B_l-\mu\right)
\end{equation}
and the integration over $y$ can be performed with the help of the property Eq.~(\ref{deltapr}) of the $\delta$-function, where 
\begin{equation}
\left\vert\frac{\rm d}{{\rm d}y}\left[B_l-\mu\right]\right\vert=\frac{4y}{\mu}\frac{1}{\sqrt{\left(1-x\right)\left(4y^2-x\right)}}\left\vert1+a_l\right\vert\, .
\end{equation}
Using Eq.~(\ref{Bl}) the evaluation of the delta function yields
\begin{eqnarray}
\sqrt{4y^2-x}&=&-l\sqrt{1-x}\pm\vert\mu\vert\\
\vert1+a_l\vert&=&\vert\mu\vert\sqrt{1-x}\, ,
\end{eqnarray}
where according to Eq.~(\ref{requir})
\begin{equation}
\mu\geq\frac{\omega}{2},\, x\leq 1
\end{equation}
has to be fulfilled. This yields for $y$
\begin{equation}
y^2_\pm=\frac{1+\mu^2}{4}\pm\frac{\mu}{2}\sqrt{1-x}\, .
\end{equation}
Furthermore, the band-overlap function Eq.~(\ref{f}) approximately equals $f_{\lambda\cdot\lambda^\prime}(\vec k,\vec q)\simeq 1$. Thus, we finally obtain for the density function of particle-hole excitations
\begin{eqnarray}
\nonumber{\rm Im}\Pi^+(\vec q,x;\mu)&=&\frac{\sqrt{3}}{4\pi}\frac{\mu}{\left(\hbar v_F\right)^2}\frac{\sqrt{x}}{\sqrt{1-x}}\Theta\left(\omega^+_\mu-\sqrt{x}\right)~~~~~\label{stauber}\\
\nonumber&\times&\left[\frac{\Theta\left(\sqrt{x}-\omega^-_\mu\right)}{y_-\sqrt{1-y_-^2}}+\frac{1}{y_+\sqrt{1-y^2_+}}\right]\\&&
\end{eqnarray}
with
\begin{eqnarray}
\omega_\mu^-&=&\theta(\mu -1)\omega^\star\\
\omega_\mu^+&=&\theta(\sqrt{3}-\mu)+\omega^\star\theta(\mu-\sqrt{3})\\
\omega^\star&=&\frac{1}{2}\sqrt{10-\mu^2-9/\mu^2}\, .
\end{eqnarray}
This result coincides with that obtained in Ref.~\cite{Stauber2010}. In the limit $\omega_q\to 0$ the intraband transitions give a nonvanishing contribution to the integral Eq.~(\ref{ImLindhard}) only for $x=\left(\frac{\omega}{\omega_q}\right)^2<1$, i.e., for small excitation energies $\omega\leq\omega_q$.
\subsection{\label{ImP-}Interband transitions for small momentum transfer}
In order to explicitely perform the limit $\omega_q\to 0$ we define $\tilde x=\frac{1}{x}=\left(\frac{\omega_q}{\omega}\right)^2<1$ and accordingly rewrite the integral Eq.~(\ref{ImLindhard}) as
\begin{eqnarray}
\nonumber&&{\rm Im}\Pi^-(\vec q,\omega;\mu)=\frac{1}{\omega}\frac{\sqrt{3}}{\pi}\frac{1}{\left(\hbar v_F\right)^2}\int_{y_{min}}^{y_{max}}\frac{{\rm d}y}{\sqrt{1-y^2}}\label{Integrand2}\\&&\nonumber
\times\frac{\Theta\left(\lambda\cdot\lambda^\prime\left[\tilde B_l-\frac{1}{2\omega_q}\right]\right)}{\sqrt{\left(1-\tilde x\right)\left(1-4y^2\tilde x\right)}}\left[\frac{T^+_{\lambda\cdot\lambda^\prime}\tilde F^+_{\lambda\cdot\lambda^\prime}}{\tilde B_+}+\frac{T^-_{\lambda\cdot\lambda^\prime}\tilde F^-_{\lambda\cdot\lambda^\prime}}{\tilde B_-}\right]\, ,\\&&
\end{eqnarray}
with
\begin{eqnarray}
\tilde a_l&=&-\cos\frac{q_x}{2}+l\sqrt{\left(1-\tilde x\right)\left(1-4y^2\tilde x\right)}\\
\tilde B_l&=&\sqrt{\tilde x}B_l\\
\tilde F^l_{\lambda\cdot\lambda^\prime}&=&\tilde xF^l_{\lambda\cdot\lambda^\prime}
\end{eqnarray}
For small values of $\tilde x$ the quantity $\tilde a_l$ can be approximated as
\begin{eqnarray}
\nonumber\tilde a^0_+&=&\lim_{\omega_q\to 0}\tilde a_+=\frac{\tilde x}{2}\left[\left(\frac{\omega}{2}\right)^2-(1+4y^2)\right]\label{notations}\\
\nonumber\tilde a^0_-&=&\lim_{\omega_q\to 0}\tilde a_-=-2+\frac{\tilde x}{2}\left[\left(\frac{\omega}{2}\right)^2+(1+4y^2)\right]\\
\nonumber b^0_+&=&\lim_{\omega_q\to 0} b_+=2\sqrt{y^2-\left\{\frac{1}{4}\left[\left(\frac{\omega}{2}\right)^2-1\right]-y^2\right\}^2}\, .
\\&&
\end{eqnarray}
There is no contribution for $l=-1$ and $\omega_q\to 0$, i.e., $\tilde x\to 0$ since the requirement Eq.~(\ref{requir}) leads to the condition $\tilde x\geq\frac{2}{\left(\frac{\omega}{2}\right)^2+1+4y^2}\geq\frac{1}{7}$ which cannot be fulfilled in that limit. Furthermore, the rule of l'H\^opital yields
\begin{equation}
\label{rat}\frac{\tilde F^+_-}{\tilde B^+_-}\simeq\omega_q^2\frac{\frac{1}{3}\left[1+2y^2\right]-\frac{1}{9}\left[\frac{\omega}{2}\right]^2-2\left[\frac{b^0_+}{\omega}\right]^2}{b^0_+}.
\end{equation}
In this case the band-overlap function Eq.~(\ref{f}) and thus the ratio Eq.~(\ref{rat}) is proportional to $\omega_q^2$. 
The DOS $\rho(\omega)$ of the tight binding model~\cite{Hobson1953} is given in terms of a complete elliptic integral~\cite{Castro2009}, 
\begin{eqnarray}
\nonumber\label{rho}\rho(\omega)&=&\frac{2\omega}{\pi^2}\int_{y_{min}}^{y_{max}}\frac{{\rm d}y}{\sqrt{1-y^2}}\frac{1}{\sqrt{y^2-\left\{\frac{1}{4}\left[\omega^2-1\right]-y^2\right\}^2}}~~~\\
y_{min}&=&\frac{1}{2}\left\vert 1-\omega\right\vert,\, y_{max}={\rm min}\left\{1,\frac{1}{2}\left(1+\omega\right)\right\}.
\end{eqnarray}
Using this relation we obtain for the case of electron-phonon excitations, where $f_{\lambda\cdot\lambda^\prime}=1$ (see Ref.~\cite{Castro2009})
\begin{equation}
\lim_{\tilde x\to 0}{\rm Im}\Pi(\vec q,\omega;\mu)=\frac{\pi\sqrt{3}}{\left(\hbar v_F\right)^2}\Theta\left(\frac{\omega}{2}-\mu\right)\rho\left(\frac{\omega}{2}\right),
\end{equation}
and for that of electron-electron polarisations considered in this paper 
\begin{eqnarray}
\label{EEDOS}
&&{\rm Im}\Pi^-(\vec q,\omega;\mu)\\&&
\nonumber\simeq\frac{\omega_q^2}{\omega}\frac{\sqrt{3}}{\pi}\frac{1}{\left(\hbar v_F\right)^2}\Theta\left(\frac{\omega}{2}-\mu\right)\\&&
\nonumber\times\int_{y_{min}}^{y_{max}}\frac{{\rm d}y}{\sqrt{1-y^2}}\frac{\frac{1}{3}\left[1+2y^2\right]-\frac{1}{9}\left[\frac{\omega}{2}\right]^2-2\left[\frac{b^0_+}{\omega}\right]^2}{b^0_+}\\&&
\nonumber=\frac{\omega_q^2}{\omega}\frac{\sqrt{3}}{\pi}\frac{1}{\left(\hbar v_F\right)^2}\Theta\left(\frac{\omega}{2}-\mu\right)\\&&
\nonumber\times\left[\frac{\pi^2}{2\omega}\left(\frac{1}{3}-\frac{1}{9}\left(\frac{\omega}{2}\right)^2\right)\rho\left(\frac{\omega}{2}\right)\right.\\&&
\nonumber\left.+\frac{2}{3}\int_{y_{min}}^{y_{max}}\frac{{\rm d}y}{\sqrt{1-y^2}}\frac{y^2}{b^0_+}
-\frac{2}{\omega^2}\int_{y_{min}}^{y_{max}}\frac{{\rm d}y}{\sqrt{1-y^2}}{b^0_+}\right]
\end{eqnarray}
with
\begin{equation}
\nonumber y_{min}=\frac{1}{2}\left\vert 1-\frac{\omega}{2}\right\vert\, y_{max}={\rm min}\left\{1,\frac{1}{2}\left(1+\frac{\omega}{2}\right)\right\}
\end{equation}
and $b^0_+$ given in Eq.~(\ref{notations}).
The limits of integration, $y_{min}$ and $y_{max}$, are obtained from the condition that the radicand of $b$ is positive. Thus ${\rm Im}\Pi^-(\vec q,\omega;\mu)$ can be expressed in terms of the DOS, i.e. an elliptic integral of the first kind~\cite{Castro2009}, one of the third kind, which both comprise a logarithmic singularity and a non-singular term. It can be readily shown, that ${\rm Im}\Pi^-(\vec q,\omega;\mu)$ drops down to zero when $\omega\geq\omega_q$ approaches zero. In fact $\frac{\omega}{\omega_q^2}{\rm Im}\Pi^-(\vec q,\omega;\mu)$ converges to $\frac{1}{4\left(\hbar v_F\right)^2}$ for $\omega\to 0$. For $\omega <\omega_q$ its contribution is negligible.
\subsection{\label{fsum}Computation of the $f$-sum rule for $\omega_q\to 0$}
The $f$-sum rule is defined as~\cite{Stauber2010}
\begin{equation}
Z(\mu)=\left(\hbar v_F\right)^2\frac{1}{4\pi}\int_0^6{\rm d}\omega\frac{\omega}{\omega_q^2}{\rm Im}\Pi(\vec q,\omega;\mu)
\end{equation}
For intraband transitions we obtain from Eq.~(\ref{stauber}) with the variable transformations $x=\left(\frac{\omega}{\omega_q}\right)^2$ yielding ${\rm d}\omega\frac{\omega}{\omega_q^2}=\frac{1}{2}dx$ and $\Omega=\sqrt{1-x}$
\begin{eqnarray}
&\label{Zp}Z^+(\mu)=\frac{\sqrt{3}}{8\pi^2}\int_0^1{\rm d}\Omega\sqrt{1-\Omega^2}\\
&\nonumber\times\left[\frac{\Theta\left(\Omega_++\Omega\right)}{\sqrt{\left(\Omega_--\Omega\right)\left(\Omega_++\Omega\right)}}+\frac{\Theta\left(\Omega_+-\Omega\right)}{\sqrt{\left(\Omega_-+\Omega\right)\left(\Omega_+-\Omega\right)}}\right]
\end{eqnarray}
with
\begin{eqnarray}
&&\Omega_-=\frac{1+\mu^2}{2\mu}\geq 1\\
&&\Omega_+=\frac{3-\mu^2}{2\mu}.
\end{eqnarray}
For $\mu=1$ we have $\Omega_-=\Omega_+=1$ and 
\begin{equation}
Z^+(\mu=1)=\frac{\sqrt{3}}{4\pi^2}.
\end{equation} 
We also computed the derivative of $Z^+(\mu)$ at $\mu=1$. With
\begin{eqnarray}
\frac{{\rm d}\Omega_+}{{\rm d}\mu}\vert_{\mu=1}&=&-\frac{3+\mu^2}{2\mu^2}\vert_{\mu =1}=-2\\
\frac{{\rm d}\Omega_-}{{\rm d}\mu}\vert_{\mu=1}&=&\frac{1-\mu^2}{2\mu^2}\vert_{\mu =1}=0
\end{eqnarray}
we obtain
\begin{eqnarray}
\label{derivIntra}
\frac{{\rm d}Z^+(\mu=1)}{{\rm d}\mu}&=&-\frac{\sqrt{3}}{4\pi^2}\int_0^1{\rm d}\Omega\\
\nonumber&\times&\left[\delta(1-\Omega)-\frac{1}{2\left(1+\Omega\right)}+\frac{1}{2\left(1-\Omega\right)}\right]
\end{eqnarray}
which is logarithmically divergent at $\mu =1$.

The $f$-sum rule for interband transitions is given with the notations Eqs.~(\ref{notations}) by
\begin{eqnarray}
\label{Zm}Z^-(\mu)=&\frac{\sqrt{3}}{4\pi^2}\int_{2\mu}^6{\rm d}\omega\int_{y_{min}}^{y_{max}}\frac{{\rm d}y}{\sqrt{1-y^2}}\\
&\times\frac{\frac{1}{3}\left(1+2y^2\right)-\frac{1}{9}\left(\frac{\omega}{2}\right)^2-2\left(\frac{2b}{\omega}\right)^2}{2b}.\nonumber
\end{eqnarray}

Its derivative with respect to $\mu$ equals its integrand evaluated at $\omega/2=\mu$, $\frac{{\rm d}Z^-(\mu=1)}{{\rm d}\mu}={\rm Im}\Pi^-(\vec q,\omega=2;\mu)$. Using that $b^+_0=y\sqrt{1-y^2}$ at $\omega =2$ yields
\begin{equation}
\label{derivInter}
\frac{{\rm d}Z^-(\mu=1)}{{\rm d}\mu}=2\rho(1)
+\frac{6}{\pi^2}\int_{0}^{1}{\rm d}y\frac{y}{1-y^2}
-\frac{18}{\pi^2}\int_{0}^{1}{\rm d}y\, y.
\end{equation}
The first and the second term are logarithmically divergent at $\mu=1$.

For small values of $\mu\ll 1$ $Z^+(\mu)$ grows limearly with $\mu$, $Z^+(\mu)\simeq\frac{1}{8\pi}\mu$, while $Z^-(\mu)$ decreases linearly, $Z^-(\mu)\simeq\frac{\pi}{24\sqrt{3}}-\frac{1}{8\pi}\mu$. Close to the $\Gamma$ point, which corresponds to $\mu=3$, $Z^-(\mu)$ is vanishingly small, $Z^-(\mu)\simeq\frac{1}{108}\frac{1}{8\pi}(3-\mu)^3$ while $Z^+(\mu)$ decreases linearly with $\mu$ approaching $\mu =3$, $Z^+(\mu)\simeq\frac{1}{8\pi}(3-\mu)$.


\begin{thebibliography}{42}
\bibitem{footnote1}
Note that the electronic energies of graphene take over the r${\rm\hat o}$le of the frequencies in the microwave Dirac billiards.
\bibitem{footnote2}
We neglect the physical spin of the electrons which would amount to a doubling of the degrees of freedom.
\end{thebibliography}

\end{document}